\documentclass[11pt, article]{revtex4}
\usepackage{amssymb}
\usepackage{latexsym}
\usepackage[dvips]{color}
\usepackage{epsfig}
\usepackage{amsfonts,amsmath}
\usepackage{latexsym}
\usepackage{graphics,color,graphicx,shortvrb}

\usepackage{textcomp}

\newtheorem{Theorem}{Theorem}[section]
\newtheorem{Definition}[Theorem]{Definition}
\newtheorem{Proposition}[Theorem]{Proposition}

\newtheorem{Minipage}{\vspace{ 5 mm}}

\definecolor{ProcessBlue}{cmyk}{1,0,0,0.25}
\definecolor{Black}{cmyk}{0,0,0,1}
\definecolor{Red}{cmyk}{0,1,1,0}
\definecolor{Green}{cmyk}{0.9,0,1,0}
\definecolor{Orange}{cmyk}{0,0.61,0.87,0.1}
\definecolor{Fuchsia}{cmyk}{0.47,0.91,0,0.06}
\definecolor{PineGreen}{cmyk}{0.92,0,0.59,0.25}

 \makeindex
\usepackage{hyperref}
\hypersetup{pdfpagemode=FullScreen}

\hypersetup{backref, hyperindex=true, citecolor=true}
       
\begin{document}

\title{Constructing and exploring wells of energy
landscapes}

\author{Jean-Pierre Aubin$^{(a)}$ and Annick Lesne$^{(b)}$}
\affiliation{(a) R\'{e}seau de
Recherche Viabilit\'{e}, Jeux, Contr\^{o}le, 14 rue Domat,  F-75005 Paris\\
J.P.Aubin@wanadoo.fr\\
(b) Laboratoire de Physique Th\'eorique des Liquides, Case 121, Universit\'e Pierre et Marie Curie,
4 place Jussieu, F-75252 Paris\\
lesne@lptl.jussieu.fr}

\begin{abstract}
\color{Black} Landscape paradigm is ubiquitous in physics and other
natural sciences,
but it has to be supplemented with both quantitative and
  qualitatively  meaningful tools for analyzing the topography
of a given landscape.
We here consider  dynamic explorations of the relief and introduce
as basic topographic features ``wells
 of duration $T$ and altitude  $y$''.
We  determine an intrinsic exploration
mechanism governing the evolutions from
 an initial state in the well up to its rim
 in a prescribed time,
whose finite-difference approximations on
finite grids yield a constructive algorithm
for determining the wells.
Our  main results are thus (i)  a quantitative
 characterization of landscape topography
rooted in a dynamic
 exploration of the landscape, (ii) an alternative
 to stochastic gradient dynamics for performing such an exploration,
 (iii) a constructive access to the wells
 and (iv) the determination of some bare
  dynamic features inherent to the landscape.

The mathematical tools  used here  are not familiar in physics: They
come from set-valued analysis (differential calculus of set-valued maps
and differential inclusions) and viability theory (capture basins of
targets under evolutionary systems) which have been developed during the
last two decades; we therefore propose
 a minimal appendix exposing them
at the end of this paper to bridge  the possible gap.

\color{Black}
\end{abstract}

\maketitle
\newpage
\color{ProcessBlue}  \section{Introduction} \color{Black}

 \color{Fuchsia}
\subsection{The landscape paradigm in natural sciences} \color{Black}

The general notion of landscape is encountered in many
different domains, for instance in
physics, neural networks (Hopfield nets, \cite[Hopfield]{hopfield}) and
learning processes, molecular biology \cite[Becker \& Karplus]{karplus}
\cite[Frauenfelder et al.]{frauen-exp},
ecology  and  evolutionary biology
\cite[Kauffman]{kauffman},
or optimization problems,
 to cite but a few.
From the mathematical viewpoint, a {\it landscape} is simply
a function  $ V:X \mapsto \mathbb{R}\cup \{+\infty\}$
(more precisely an extended \footnote{We call $V$ an ``extended
function" in the sense that it can take the value $+\infty$. This allows
us to encapsulate state constraints $x\in K$
 (where $K$ is the  subset of  ``viable" (acceptable, feasible)
 states) in the definition of the extended function,
 equal to $+\infty$
outside $K$. A well-known example in physics is hard-core interaction
potential.}
function since it might takes infinite values $+\infty$)
associating a  real value $V(x)$
to each state $x\in X$ of the system.
>From a physical viewpoint,
the  status and definition  of $V$
strongly depend on the scale at which the system is described,
 reflecting in the choice of the space of states $X$.

Let us give some examples to sustain our exposition.
 In statistical physics and molecular biology,
 $V(x)$ can be the {\it energy landscape} if $x$ is the
 (high dimensional) microscopic configuration of the
considered system: atomic coordinates in a
glass \cite[De Benedetti \& Stillinger]{stillinger-nature},
 spin orientations
in a spin glass \cite[Fischer \& Hertz]{hertz}, tridimensional
conformation of the hundred or more amino-acids forming a
protein \cite[Frauenfelder]{frauen}, spatial positions of bead centers in a
granular medium \cite[Edwards]{edwards}. It can also be a
(mesoscopic) {\it free energy
landscape}  if
$x$ is the value of a (low dimensional)
 order parameter describing the global state of the system:
spatially average density, overall magnetization, conformational parameter(s)
for a macromolecule (as for instance its radius of gyration).
At a still more macroscopic level, $x$  can be
 a reaction coordinate
measuring the progress along a path representing some transformation
 of the system
and inscribed on an {\it effective energy landscape}.
In quite different contexts, cost functions encountered
in optimization problems are close analogues to energy landscapes
\cite[Attouch \& Soubeyran]{as04landsat},
whereas fitness landscapes encountered in ecology and evolutionary
biology can be cast in the frame of free or effective energy
 landscapes, up to a sign change (namely, by considering the
 opposite of the fitness).
 (See \cite[Sherrington]{Sherrington} for
 an introductory review).

Energy or free energy landscapes are currently exploited in
 stochastic
gradient methods accounting for the interplay between thermal motion and
interaction forces (effective forces in the case of a free energy
landscape) deriving from the potential $x \mapsto V(x)$.
In complex systems (glasses, spin glasses, proteins, for instance)
the landscape $V$ typically presents
 a large number of local optima around which the
solution of
 a stochastic gradient
method is trapped and travels a long time  before going away and visit
other local minima. This dynamical behavior has  been advocated by
Giorgio Parisi to encapsulate a meaning of complexity and rugged
landscapes are often seen as a mark of complex systems. (See for instance
\cite[Mezard, Parisi \& Virasoro]{mezard}.)

Although the landscape $V(x)$ is thus endowed with
 different status and interpretations in varying contexts,
understanding and controlling the system behavior requires in any case
a quantitative knowledge of the landscape topography.
 It is thus of
the utmost importance to design efficient tools allowing a dynamical
analysis of local minima of such a function $x \mapsto V(x)$.
We emphasize  that it is not just an academic issue since actual energy or
free energy landscapes of real systems are available through either:
\begin{itemize}  \item  a \textit{theoretical access} from first principles (e.g. molecular
interactions, spin-spin interactions) and/or modeling hypotheses,
allowing to write an explicit formula for $V(x)$;
 \item  an \textit{experimental access}, for instance for proteins (indirect
 kinetic or spectral measurements)
 \cite[Frauenfelder {\it et al.}]{frauen-exp};
  \item a \textit{numerical access}, either through molecular dynamics at
 atomic scale, yielding the energy landscape,
either through Monte Carlo sampling of the configuration space
 according to the Boltzmann distribution,
  yielding free energy landscapes for the relevant order
parameter(s) of the system \cite[Frenkel \& Smit]{frenkel}.
     \end{itemize}

\color{Fuchsia}
\subsection{Dynamical analysis of a landscape topography} \color{Black}

 We here propose a theoretical and algorithmic analysis allowing us to
determine
 quantitatively the landscape relief of a function $V$, e.g.
location of wells, location and heights of the barriers associated with a
given dynamics for exploring the landscape
\footnote{
Care has to be taken  not to confuse
exploration mechanism (valid whatever the interpretation of $V$ is)
and actual evolution (whose modeling is indissociable from
the status and definition of $V$). We here mainly consider
exploration dynamics.
Only at the end of the paper, it is suggested that the intrinsic exploration
mechanism determined in Section \ref{sec-3} could be a relevant alternative
to stochastic gradient dynamics for modeling actual dynamic
features of the system.}.
 It gives
access to a hierarchical picture of the landscape and allows us
 to determine the
nesting of wells and barriers at different scales.

Given a dynamic exploration mechanism (such as a stochastic gradient
dynamics), we define  the ``wells of velocity $\lambda$, duration $T$ and
altitude $y$'' as the sets of initial states
$x\in X$  ``below the level $y$'', i.e. $V(x)\leq y$,
from which at least one evolution governed by the exploration mechanism,
and of velocity bounded by $\lambda$,
reaches the rim $y$ of the well at the prescribed time $T$. When the well
is not empty, we
then evidence
intrinsic dynamics  governing the evolutions from an initial
state of the well up to its rim $y$
 at prescribed time $T$. This intrinsic exploration
mechanism is characterized from \textit{the time-derivative of the well,
regarded as a set-valued map associating with the prescribed duration $T$ and
the altitude $y$ the elements of the well}.
Both the wells and their intrinsic exploration mechanism can be approximated by
finite-difference approximations on finite grids,
{\it which allows to implement a constructive algorithm}.

This study offers an alternative to
 stochastic gradient-type exploration mechanisms.
In quite a similar way of thought,
second-order exploration mechanisms of the graph of an energy landscape
function has been proposed in \cite[Attouch, Goudou \& Redont]{agr00hbal}.
 We here suggest to start
the landscape exploration
  with a universal mechanism, independent of the energy function,
allowing us to look at any possible velocity with prescribed norm
$\lambda$ and to
retain its intrinsic exploration dynamics as a good candidate for a dynamical
system exploring the given energy landscape.
The stochastic gradient method is thus
replaced by a differential inclusion involving the time-derivative of the
well, but allowing in the same spirit the system state to escape the trap
of local minima, while being quantitatively influenced by their depth.

The mathematical tools we use  are quite novel in physics: They
come from set-valued analysis (differential calculus of set-valued maps
and differential inclusions) and viability theory (capture basins of
targets under evolutionary systems) which have been developed
during  the last two
decades.

\vskip 3mm
The resulting quantitative topographic description by  wells rooted in a
constructive dynamic exploration of the landscape and the associated
determination of the statistical properties of its relief can then be
exploited for:

\begin{enumerate}

  \item  \textit{performing a quantitative
characterization of the landscape},
 for comparison or classification purposes. It allows to
investigate bifurcations, more currently called
 phase transitions in many-particle systems
  \cite[Lagu\"es \& Lesne]{llbook};

\item  \textit{providing a
quantitative access to the landscape hierarchical structure} and allowing to
estimate its ruggedness, which yields a tentative measure of the system
complexity;

 \item \textit{defining macrostates and macroscopic variables}
to be used in coarse-grained descriptions of the system. The relevance of
such an approach is to provide an intrinsic determination of macrostates,
founded upon the identification of macroscopic features with slow modes
and slowly evolving properties \cite[Gaveau, Lesne \& Schulman]{GLS}.
 \end{enumerate}

\vskip 3mm

{\bf Outline of the paper:} --- \hspace{ 2 mm}
 In  Section~II, we shall
define wells, introduce some  mathematical features of their relief, and
reformulate their characterization in terms of ``capture basin of a
target'', a key concept of viability theory which finds here a unexpected,
yet natural,  application. In Section~III, we present the algorithm allowing
to construct explicitly these wells
and the intrinsic exploration mechanism on which it is based.
 In Section~IV,
 we introduce the notion of complete wells, matching more closely
with physical landscape features. After a conclusive summary in Section~V,
the essential notions of viability theory needed for this paper are
presented in an Appendix.
 $\;\;
\blacksquare$ \normalsize \vspace{ 5 mm}

\color{Black}

 \color{ProcessBlue}
\section{Wells of an energy landscape} \color{Black}

\color{Fuchsia}
\subsection{An efficient alternative to stochastic exploration} \color{Black}
In order to provide both a quantitatively meaningful and
 quantitative topographic
analysis of a landscape $V$ on a space $X$,
we introduce  ``wells of duration $T$
 and depth $y$'' .
Given a  dynamical system,
allowing upwards steps of velocity
  bounded by a parameter $\lambda$,  these wells are the sets
${\bf P}_{V}(\lambda;t,y)$ of initial states ``below the
 level $y$'' (i.e. of states $x\in X$
such that $V(x)\leq y$)
from which at least one \footnote{or, if needed, all
 evolutions.} evolution
 reaches the upper level $y$ (what we call the 
{\it rim of the well}) at time $T$.
In other words, given some tolerance $\lambda$ allowing upwards
exploration, and some level $y$, the wells and their depth might be
dynamically (the experimentally meaningful and operational way) determined
according to the trapping time $T$.

For exploratory purposes, we
here implement an alternative to
 stochastic gradient dynamics
and replace   stochastic differential
 equations
encountered in physics \color{Black} by differential inclusions of the
form \footnote{At odds with stochastic models currently used in physics,
differential inclusions provide another way
to translate uncertainty in mathematical terms.
 A full corpus of results have been steadily  accumulated since
their introduction in the early 1930's by Marchaud and Zaremba, and next,
by the Polish and Russian schools, around Wa\.zewski and Filippov, who
laid the foundations of the mathematical theory of differential inclusions
after the 1950's. The confrontation of evolutions governed by differential
inclusion and viability constraints for reaching targets in finite time
began in the early 1980's to provide regulation maps and feedbacks. Their
use for translating mathematically  uncertainty ``against nature'' as an
alternative to stochastic differential equations  began at the same
period, but is taking a new start when it was realized recently that they
yield much more general results than the stochastic paradigm: Differential
inclusions allow us to represent uncertainty by state-dependent maps. As a
name is needed to refer to such a viewpoint and to underline its
difference with stochastic differential equations, we called these {\it
tychastic systems} when the velocities depend upon both the state and a
parameter that is called  ``tyche'' (meaning
``chance'' in classical Greek, from the Goddess Tyche of (good  and bad)
fortune), which play the role of perturbations or disturbances on
 which actors and  decision makers have no control.
Tyches could in fact have been more plainly
 called ``random variables'' if this
vocabulary were not already confiscated by probabilists. This is why we
borrow   the term of {\em tychastic evolution\/} to the American
philosopher Charles Peirce who introduced it in a paper published in 1893
under the title \textit{``Evolutionary love''}: {\em ``Three modes of
evolution have thus been brought before us: evolution by  fortuitous
variation, evolution by mechanical necessity, and evolution by creative
love. We may term them tychastic evolution, or tychasm, anancastic
evolution, or  anancasm, and agapastic evolution, or agapasm. \/}}
\begin{displaymath} \forall \; t \geq 0, \;  \; x' (t) \; \in  \;
     F(\lambda;x(t))
 \end{displaymath}
where $x\leadsto F(\lambda;x(t))$ is some set-valued function on $X$
(i.e. $F(\lambda;x)$ is a subset of $X$)
parameterized by a parameter  $\lambda \in  \mathbb{R}^{}$.
Compared to a differential equation, the solution of a differential inclusion
is less constrained since the full
 specification of the derivative
$x'(t)$ at each time $t$ is replaced by a constraint
on the region $F(x(t))\subset X$ where it has to
 lie.
Such a tolerance is highly valuable and quite realistic in the modeling of
an actual system, since the experimentally available knowledge about its
dynamics generally provides only bounds
 (or more generally viability constraints) on the kinetic rates, rather
than explicit pointwise expressions of these rates as a
 function of the system state.
These bounds might nevertheless vary with the system state $x(t)$,
hence defining a specific set $F(x(t))$ at each time $t$.
For instance, in the case when the
function $V$ is differentiable,
a close analog to stochastic gradient
dynamics is provided by

\begin{displaymath}
 x'(t) \; \in \; - \nabla V(x(t)) + \lambda B
\end{displaymath}
where $B$ denotes the unit ball of the finite-dimensional vector space
$X$. Indeed, the gradient dynamics $x'(t) = - \nabla V(x(t))$ governs
evolutions decreasing along the function $V$, but stopping at the first
encountered  local minimum. To overcome this stalling situation, a natural
idea is to perturb the gradient equation either by a stochastic noise as
currently implemented
in simulated annealing methods or, as we suggest here, by a ``{\it tychastic}''
one. Indeed, differential inclusion $x'(t) \in - \nabla V(x(t)) + \lambda
B$ is the ``tychastic version'' of the stochastic differential equation
$dx=- \nabla V(x(t))dt +\lambda dW(t)$  (see \cite[Aubin \&
Doss]{ad01stoc} for the links between stochastic and tychastic viability).

However, we  have to overcome the fact that the function $V$ is usually
not differentiable, if obtained through experimental measures or
simulations
 and no longer analytically defined.
Hence the concept of gradient disappears (when the observable or simulated
configuration space is discrete), or has to be approximated by gradients
of functions interpolating in one way or another the experimental data.
Any method allowing to bypass these obstacles and to deal with graphs  of
such functions may be worth of being investigated.

Another suggestion is to leave open the choice of the directions of
exploration  by looking for any way to climb the landscape
$V$ to reach a given level $y$ at a given time $T$. For that purpose, we
can choose  $F(\lambda;x):= \lambda B$, stating that any velocity of norm
$\lambda$ is \textit{a priori} an eligible candidate to apply for such a
mission. We shall provide below
 the way of further  selecting the most efficient (subset of)
velocity(ies), i.e. achieving the most thoroughly and the most efficiently,
from a numerical viewpoint, the quantitative
 exploration of the landscape relief.
The same type of strategy has been use in previous works for constructing
an algorithm
 that is also of relevance for
landscapes. This so-called
 Montagnes Russes Algorithm converges to global minima of
an extended function jumping over local minima, which amounts to use the
gradient algorithm to the  smallest of the exponential Lyapunov functions
above the energy function for the differential inclusion $x'(t) \in
\lambda B$ (See \cite[Aubin \& Najman]{an94mr,an94rm})). But whereas this
algorithm was devoted to the search of global minima, we are here looking
for exploratory tools providing a complete hierarchical picture of the
landscape.

\color{Fuchsia}  \subsection{Definition and characterization of wells}
\color{Black}

 From now on, we assume that
the set-valued map $x  \leadsto F(\lambda;x)$ governing the exploration
dynamics is given. We denote by $y \in  \mathbb{R}^{}$ the altitude of the
well we wish to study.
$y=0$ is set through the (arbitrary) choice of a base level
 (or, if known and finite, by the lower bound on $V$).
 Usually, the relevant altitudes are the values of
the {\it local maxima} or {\it saddle points}
 of the function $V$. We shall associate with it
 the concept of well ${\bf P}(\lambda; T,y)$ of duration $T$
 and altitude $y $ defined as follows:

\vspace{ 5 mm} \fbox{\begin{minipage}{14.0cm} \begin{Minipage}
\label{mpdef:well}
\begin{center} \color{Red} {\large \bf Wells of a function
 under a differential inclusion} \end{center}  \end{Minipage}
\footnotesize \color{PineGreen}
\begin{Definition}\label{def:well} Consider an extended
function $ V:X \mapsto \mathbb{R}\cup \{+\infty\}$ and a differential
inclusion
\begin{displaymath} \forall \; t \geq 0, \;  \; x' (t) \; \in  \;
        F(\lambda;x(t))
 \end{displaymath}
Denote by
 \begin{displaymath}
  {\bf S} (V, y) \; := \; \{x \in X \;
\mbox{ such that} \; V(x) \; \leq \; y\} \; \; and \; \;  {\bf S}_{0} (V,
y) \; := \; \{x \in X \; \mbox{ such that} \; V(x) \; = \; y\}
\end{displaymath}
the \index{level set} \textsf{level sets} of the function $V$ and by
${\cal S}_{\lambda}(x)$ the set of solutions to the above differential
inclusion starting at $x$. The \index{well of a function} \textsf{well}
${\bf P}_{V}(\lambda; T,y) \subset {\bf S}(V;y)$ of duration $T$
and altitude $y $ of the function $V$ is defined by the set of initial
states $x \in {\bf S}(V;y)$ such that there exists at least one solution
$x_{\lambda}(\cdot) \in {\cal S}_{\lambda}(x)$ such that

\begin{displaymath} \left\{ \begin{array}{ll}
 (i) & V (x_{\lambda}(T)) \; = \; y\\
  (ii) & \forall \; t \in [0,T], \; \; V (x_{\lambda}(t)) \; \leq \; y \end{array} \right.
\end{displaymath}
  \end{Definition}
\color{Black} \end{minipage}} \normalsize \vspace{5 mm}

We observe that

${\bf P}_{V}(\lambda;0,y) \; = \; {\bf S}_{0} (V, y) \; := \; \left\{ x \in
X \; \mbox{ such that} \;V (x) \; = \; y\right\}$.
In other words, the well ${\bf P}_{V}(\lambda; T,y)$ is the set of initial
conditions $x$ in the well from which there exists at least one evolution
$x_{\lambda} (\cdot)$ staying below
the level $y$ during a duration  $T$ and
reaching the level $y$ at exactly time $T$. This does not exclude the fact
that for some earlier time $t^{\star} \leq T$ (or  some later time
$t^{\star} \geq T$), the evolution reaches the level $y$. This just means
that  $x$  belongs to the intersection ${\bf P}_{V}(\lambda; T,y) \cap
{\bf P}_{V}(\lambda; t^{\star},y)$ of wells of several durations.
This point can be made more explicit:
considering the initial state $x$
 of the system as a variable and the time to reach the level
$y$ as the result, we can define
 the \textsf{reaching function} $(\lambda,x,y) \mapsto \xi
(\lambda,x,y)$ by

\begin{displaymath}
\xi (\lambda,x,y) \; := \; \inf _{x \in {\bf P}_{V}(\lambda; T,y)}T
\end{displaymath}
providing the first instant when one evolution starting from $x$ reaches
the level $y$.

We can also regard the same object by introducing the set-valued map
$(\lambda;T,x)  \leadsto {\bf P} ^{-1}_{V}(\lambda; T,x) $ associating
with the parameter $\lambda$, the duration $T$ and the initial state $x$
the altitude $y$ of the well the rim of which can be reached at time $T$
by at least an evolution governed by differential inclusion $x' (t) \in
F(\lambda;x(t))$.

Turning back to the initial definition, the
{\it maximal depth} $\delta_{V}(\lambda; T,y)$
of the well ${\bf P}_{V}(\lambda; T,y)$ is
defined by
$$\delta_{V}(\lambda; T,y)
\; := \; \sup _{x \in {\bf P}_{V}(\lambda; T,y)}(y - V(x))$$
The knowledge of the wells provide some physical characteristics of the
landscape $V$,  thus bridging the above mathematical definitions with a more
traditional description of landscapes.
 We observe for instance  that $\xi (\lambda,x,y)$ is the \textsf{escape time}
for the given dynamics,
also called the \textsf{first passage time}, from above a barrier of top $y$
when the velocity is bounded by $\lambda$.
Its inverse  $[\xi(\lambda,x,y)]^{-1}$
 has the meaning of a kinetic constant.

Denoting
$\Omega _{V}(\lambda; T,y)$  the number of  the connected components of
well ${\bf P}_{V}(\lambda; T,y)$, its logarithm
 is the\index{configurational entropy} \textsf{configurational entropy}
 (See \cite[Stillinger]{stillinger-PRE} and
\cite[Edwards]{edwards} for its meaning and use in physics, respectively
for glasses and granular media):

\begin{displaymath}
\sigma _{V}(\lambda; T,y) \; := \; \log \left(\Omega_{V}(\lambda; T,y)
\right)
\end{displaymath}

\vskip 3mm \color{Black} In summary, what
 we are basically looking for is the
subset of $(x,y,\lambda,T)$ such that either $x \in {\bf P}_{V}(\lambda;
T,y)$, or $T \geq \xi (\lambda,x,y)$ or $y \in {\bf P}_{V} ^{-1}(\lambda;
T,x)$. As detailed in the next section, we shall give
a mathematical characterization of this set as a ``capture basin of a
target under an auxiliary system'',
allowing to implement a constructive algorithm.
  \color{Black} We choose here the
representation  of this set  through the above concept of well $x \in {\bf
P}_{V}(\lambda; T,y)$.

\subsection{Viability characterization of wells}

The next step of our investigation is to translate the above
topographically meaningful features in terms of  capture basins for which many
properties have been established and constructive algorithms are available
(See the Appendix and for further details,
\cite[Aubin]{avt,a00capb,a00hjbi}) and \cite[Aubin, Bayen, Bonneuil \&
Saint-Pierre]{ab2sp00}).

\vspace{ 5 mm}\fbox{\begin{minipage}{14.0cm}
\begin{Minipage} \label{mpprp:viabcahrwell}
\begin{center}\color{Red} {\large\bf Viability characterization of
wells}\color{Black}\end{center}  \end{Minipage} \footnotesize
\begin{Proposition} \label{prp:viabcahrwell}
Consider an extended function $ V:X \mapsto \mathbb{R}\cup
\{+\infty\}$ and a differential inclusion
\begin{displaymath} \forall \; t \geq 0, \;  \; x' (t) \; \in  \;
        F(\lambda;x(t))
 \end{displaymath}
We associate with it the auxiliary system of differential inclusions

\begin{equation} \label{eq:welldifcinclaux} \left\{ \begin{array}{ll}
 (i) & x'(t) \; \in \; F(\lambda(t);x(t))\\
(ii) & y'(t) \; = \; 0\\
 (iii) & \lambda'(t) \; = \; 0\\
 (iv) & \tau'(t) \; = \; -1\\
 \end{array} \right. \end{equation}
the constrained set ${\cal K}$ and the target ${\cal C}$ defined by

\begin{displaymath}
{\cal K} \; := \; {\cal E}p(V) \times \mathbb{R}_{+}\times \mathbb{R}_{+}
\; \; and \; \; {\cal C} \; := \; \mbox{\rm Graph}(V) \times
\mathbb{R}_{+} \times \{0\}
\end{displaymath}
where $\mbox{\rm Graph}(V)$ and $ {\cal E}p(V)\subset X\times
\mathbb{R}\cup \{+\infty\}$ are respectively
the graph and epigraph of $V$ (see the Appendix for a precise definition).
Then

\begin{displaymath}
{\bf P}_{V}(\lambda; T,y) \; = \;  \left\{ x \in X \; \mbox{ such that} \;
(x,y,\lambda,T) \; \in \; \mbox{\rm
Capt}_{(\ref{eq:welldifcinclaux})}({\cal K},{\cal C}) \right\}
\end{displaymath}
where ${\rm Capt}_{(\ref{eq:welldifcinclaux})}({\cal K},{\cal C})$ is the
capture basin of the target ${\cal C}$ under evolutionary system
(\ref{eq:welldifcinclaux}) and
 under the constraint of remaining
in ${\cal K}$ (see Definition~\ref{def:capturebasin} below).
\color{Black}

\end{Proposition} \color{Black}\end{minipage}}
\normalsize \vspace{ 5 mm}

\footnotesize {\bf Proof} --- \hspace{ 2 mm} Indeed, to say that
$(x,y,\lambda,T) \; \in \; \mbox{\rm
Capt}_{(\ref{eq:welldifcinclaux})}({\cal K},{\cal C})$ amounts to saying
that there exist one evolution $x_{\lambda}(\cdot) \in {\cal
S}_{\lambda}(x)$ and a time $t^{\star} \geq 0$ such that the associated
auxiliary  evolution $$t  \to (x(t), y(t); \lambda(t); \tau(t)) \; =
\; (x(t), y,\lambda,  T-t)$$ starting from $(x,y,\lambda,T)$ at $t=0$
reaches the target ${\cal C}$ at time  $t^{\star}$ while staying meanwhile
in ${\cal K}$ :
\begin{displaymath} \left\{ \begin{array}{ll}
(i) & (x(t^{\star}),y, \lambda,T-t^{\star}) \; \in \; {\cal C}\\
(ii) & \forall \; t \in [0,t^{\star}], \; \; (x(t),\lambda,y,T-t) \; \in
\; {\cal K}\\
\end{array} \right. \end{displaymath}
The first condition is equivalent to both equations $t^{\star}= T$ and
$V(x(T))=y$. The second equation means that for every $t \in [0,T]$,
$V(x(t)) \leq y$. These are the very properties stating that $x$ belongs
to the well ${\bf P}_{V}(\lambda; T,y) $, or, equivalently, that
$\xi(\lambda;x,y) \leq T$.
 $\;\; \blacksquare$
\normalsize \vspace{ 5 mm}

Therefore, the graph of the set-valued map $(\lambda,T,y) \leadsto {\bf
P}_{V}(\lambda; T,y)$ inherits the properties of capture basins.
For instance, it can be shown
(using Theorem~\ref{captvkthm002} given in the Appendix)
 that the well satisfies a
kind of dynamical programming principle which can
be stated in the following
way:

\vspace{ 5 mm}\fbox{\begin{minipage}{14.0cm} \begin{Minipage} \label{}
\begin{center}\color{Red} {\large\bf Tracking property}\end{center}  \end{Minipage}
\footnotesize \color{Black} \begin{Proposition} The set-valued map ${\bf
P}_{V}$ is the \textsf{unique} set-valued map $(\lambda,T,y) \leadsto {\bf
P}(\lambda; T,y)$ satisfying the initial condition
\begin{displaymath}
{\bf P}(\lambda;0,y) \; := \; {\bf S}_{0} (V, y) \; := \; \left\{ x \in X
\; \mbox{ such that} \;V (x) \; = \; y\right\}
\end{displaymath}
the constraints
\begin{displaymath}
{\bf P}(\lambda; T,y) \subset {\bf S}(V;y)
\end{displaymath}
and the ``tracking property'': for any $x \in {\bf P}(\lambda; T,y) $, any
evolution $x_{\lambda} (\cdot) \in {\cal S}_{\lambda}(x)$ starting from
$x$ at time $0$ climbing the well until it reaches the rim at time $T$
satisfies

\begin{displaymath} \left\{ \begin{array}{ll}
(i) & \forall \; t \in [0,T], \; \; x(t) \; \in \; {\bf P}(\lambda;
T-t,y)\\
(ii) &  \forall \; s\geq T \; \mbox{ such that} \; \forall \; t \in [T,s],
\;\; V(x(t)) \leq y,
\;\; {\rm then}\;\; x(t) \; \in \; {\bf P}(\lambda; t-T,y)\\
 \end{array} \right.
\end{displaymath}
\end{Proposition}
\color{Black}\end{minipage}} \normalsize \vspace{ 5 mm}

\color{Fuchsia}  \subsection{Time derivative of the well as an  \textit{a
posteriori} exploratory dynamical system} \color{Black}

Since we have related the well of a landscape function to capture basins,
the basic viability theorems provide tangential characterization of the
wells allowing us to find the underlying dynamical system governing the
evolutions of differential inclusion climbing the wells up to their rims.
This can be done to the price of using
 differential calculus of set-valued maps (invented in the beginning of
 the 1980's for this purpose): Knowing the ``derivatives'' with respect to
time  of the set-valued map  $t  \leadsto {\bf P}_{V}(\lambda; t,y)$ (see
Definition~\ref{def:contingentderivative} for a rigorous definition), we
obtain an intrinsic exploration mechanism of the well:

\vspace{ 5 mm}\fbox{\begin{minipage}{14.0cm} \begin{Minipage} \label{}
\begin{center}\color{Red} {\large\bf The intrinsic exploration mechanism}\end{center}  \end{Minipage}
\footnotesize \color{Black} \begin{Proposition} For any $x \in {\bf
P}_{V}(\lambda; T,y)$, those evolutions $x_{\lambda}(\cdot) \in  {\cal
S}_{\lambda}(x)$ starting at $x$ and climbing the well ${\bf
P}_{V}(\lambda; T,y)$ in the sense that $V(x_{\lambda}(t)) \leq y $
for any $t\in[0,T]$ and
$V(x_{\lambda}(T))= y $ are governed by differential inclusion
\begin{displaymath}
x'(t) \; \in \;  -\frac{\partial {\bf P}_{V}(\lambda;T-t,y)}{\partial t}
\cap F(\lambda;x(t))
\end{displaymath}
In particular, taking for initial exploration mechanism the set-valued map
$F(\lambda;x):= \lambda B$ independent of the energy function $V$ instead
of exploration mechanisms  $ F(\lambda;x):= - \nabla V(x(t)) + \lambda B$
already dependent of $V$, we obtain a more intrinsic  exploration
mechanism.
\end{Proposition}
\color{Black}\end{minipage}} \normalsize \vspace{ 5 mm}

Theorem~\ref{thm:pdiwell} stated in the Appendix
 gives a technically precise meaning to
this symbolic statement.
In other words, the underlying dynamical system governing the evolutions
 climbing the wells up to their rims is the set
of velocities $v \in F(\lambda;x)$ pointing to  the time derivative of the
well in order to climb it from $-T$ to $0$ in order to reach the rim of
the well at altitude $y$.
The associated mathematical problem to comfort this intuitive result
starts with the definition of the time derivative and the proof of this
result is  based on results of viability theory.
Let us just mention the informal version of
Theorem~\ref{thm:pdiwell} stated in the Appendix:

\vspace{ 5 mm}\fbox{\begin{minipage}{14.0cm} \begin{Minipage} \label{}
\begin{center}\color{Red} {\large\bf Well as a solution to a partial
differential inclusion}\end{center}  \end{Minipage} \footnotesize
\color{Black} \begin{Proposition}
 The set-valued map ${\bf
P}_{V}$ is the \textsf{unique} ``Frankowska solution to the partial
differential inclusion''
\begin{displaymath}
\forall \;  t>0, \; x \in {\bf P}(\lambda; T,y),\; \;0 \; \in \;
\frac{\partial {\bf P}(\lambda; T,y)}{\partial t} + F(\lambda;x)
\end{displaymath}
satisfying the initial condition

\begin{displaymath}
{\bf P}(\lambda;0,y) \; := \; {\bf S}_{0} (V, y) \; := \; \left\{ x \in X
\; \mbox{ such that} \;V (x) \; = \; y\right\}
\end{displaymath}
and the constraints
\begin{displaymath}
{\bf P}(\lambda; T,y) \subset {\bf S}(V;y)
\end{displaymath}\end{Proposition}
\color{Black}\end{minipage}} \normalsize \vspace{ 5 mm}

 We propose now to check the
same statement in the discrete case, which allows us to define an
algorithm providing the wells under discrete dynamics and the exploratory
mechanisms.

\color{ProcessBlue}  \section{The Saint-Pierre Capture
Basin Algorithm}\label{sec-3}
\color{Black}
 The Saint-Pierre Capture
Basin Algorithm provides both the set-valued map ${\bf P}_{V}$ and for any
$x \in {\bf P}_{V}(\lambda; T,y)$, the evolutions climbing the well up to
its rim under  given duration.

Let us consider  any discrete time approximation $\Phi(\lambda ;x)$ of
$F(\lambda;x)$ governing the evolution of  sequences $\overrightarrow{x}
\in  \overrightarrow{{\cal S}}_{\lambda}(x)$ governed by

\begin{displaymath}
x_{n+1} \; \in \;  \Phi(\lambda;x_{n})
\end{displaymath}
(For instance, $\Phi(x):x + h F_{h}(\lambda;x)$ where $h$ is a time step
and $F_{h}$ an approximation of $F$ in the sense that the graph of $F_{h}$
converges to the graph of $F$ in the Painlev\'e-Kuratowski sense). The
discrete version of a well defined by Definition~\ref{def:well} for
continuous time systems becomes:

\vspace{ 5 mm} \fbox{\begin{minipage}{14.0cm} \begin{Minipage} \label{}
\begin{center} \color{Red} {\large \bf Discrete wells of a function under
a set-valued map} \end{center}  \end{Minipage} \footnotesize
\color{PineGreen}
\begin{Definition}

 Consider an extended function $ V:X \mapsto
\mathbb{R}\cup \{+\infty\}$ and a set-valued map $(\lambda,x) \leadsto
\Phi(\lambda ;x)$. The \index{discrete time well}\textsf{discrete time
well} $\overrightarrow{{\bf P}}_{V}(\lambda; N,y)\subset {\bf S}(V;y)$ of
duration $T$ and depth $y $ of the function $V$ is the
subset  of initial states $x \in {\bf S}(V;y)$ such that there exists one
sequence $\overrightarrow{x} \in {\cal S}_{\lambda}(x)$ such that

\begin{displaymath} \left\{ \begin{array}{ll}
 (i) & V (x_{N}) \; = \; y\\
  (ii) & \forall \; n \in \{0,N\}, \; \; V (x_{n}) \; \leq \; y \end{array} \right.
\end{displaymath}
\end{Definition}
\color{Black} \end{minipage}} \normalsize \vspace{5 mm}

In the discrete time, we obtain the intrinsic exploration mechanism under mere
inspection:

\vspace{ 5 mm}\fbox{\begin{minipage}{14.0cm} \begin{Minipage} \label{}
\begin{center}\color{Red} {\large\bf The discrete intrinsic
exploration mechanism}\end{center}  \end{Minipage}
\footnotesize \color{Black} \begin{Proposition} Knowing the well
$\overrightarrow{{\bf P}}_{V}$, the discrete dynamical system
$$x _{n+1} \; \in \; \Phi(\lambda;x_{n}) \cap\overrightarrow{{\bf
P}}_{V}(\lambda; N-n,y)$$ governs the evolutions starting from $x \in
\overrightarrow{{\bf P}}_{V}(\lambda; N,y)$ and arriving at step $N$ at
some $x_{N} \in \overrightarrow{{\bf P}}_{V}(\lambda;0,y)={\bf S}_{0} (V,
y)$ of the rim of the well $\overrightarrow{{\bf P}}_{V}(\lambda; N,y)$.
\end{Proposition}
\color{Black}\end{minipage}} \normalsize \vspace{ 5 mm}

In the discrete case, the discrete well is obtained by the Capture Basin
Algorithm:

\vspace{ 5 mm}\fbox{\begin{minipage}{14.0cm} \begin{Minipage} \label{}
\begin{center}\color{Red} {\large\bf Saint-Pierre Capture
Basin Algorithm}\end{center}  \end{Minipage} \footnotesize \color{Black}
\begin{Proposition} The Saint-Pierre Capture Basin Algorithm yields the
discrete well as the intersection of the following subsets defined
recursively by

\begin{displaymath} \left\{ \begin{array}{ll}
(i) & \overrightarrow{{\bf P}}_{V}(\lambda;0 ,y) \; = \;   {\bf S}_{0} (V, y)\\
(ii) & \forall \; N \geq 0, \; \; \overrightarrow{{\bf P}}_{V}(\lambda;
N+1,y) \; = \; \Phi(\lambda;\cdot)^{-1} (\overrightarrow{{\bf
P}}_{{\bf  v}}(\lambda; N,y)) \cap   {\bf S} (V, y) )\\
 \end{array} \right. \end{displaymath}

When $\Phi (\lambda;x):= x+ \lambda B$, this algorithm can be written

\begin{displaymath} \left\{ \begin{array}{ll}
(i) & \overrightarrow{{\bf P}}_{V}(\lambda;0 ,y) \; = \;   {\bf S}_{0} (V, y)\\
(ii) & \forall \; N \geq 0, \; \; \overrightarrow{{\bf P}}_{V}(\lambda;
N+1,y) \; = \;  (\overrightarrow{{\bf
P}}_{{\bf  v}}(\lambda; N,y)+\lambda B) \cap   {\bf S} (V, y) \\
 \end{array} \right. \end{displaymath}

\end{Proposition}
\color{Black}\end{minipage}} \normalsize \vspace{ 5 mm}

\footnotesize {\bf Proof} --- \hspace{ 2 mm} Indeed, we introduce the
auxiliary system $\Psi $ by

\begin{displaymath}
\Psi(x,y,\lambda, \tau) \; := \; \Phi(\lambda,x) \times \{y\} \times
\{\lambda\} \times \{\tau -1\}
\end{displaymath}
governing the evolution of the sequence
\begin{equation} \label{eq:welldifcinclauxdis} \left\{ \begin{array}{ll}
 (i) & x_{n+1} \; \in \; \Phi(\lambda;x_{n})\\
 (ii) & y_{n+1} \; = \; y_{n}\\
 (iii) & \lambda_{n+1} \; = \; \lambda_{n}\\
(iv) & \tau_{n+1}\; = \; \tau _{n}-1\\
  \end{array} \right. \end{equation}
and the constrained set ${\cal K}$ and the target ${\cal C}$ defined by

\begin{displaymath}
{\cal K} \; := \; {\cal E}p(V) \times \mathbb{R}_{+}\times \mathbb{R}_{+}
\; \;{\rm  and}
 \; \; {\cal C} \; := \; \mbox{\rm Graph}(V) \times \mathbb{R}_{+}
\times \{0\}
\end{displaymath}
Then one can prove as in the continuous time case that

\begin{displaymath}
\overrightarrow{{\bf P}}_{V}(\lambda; N,y) \; = \;  \left\{ x \in X \;
\mbox{ such that} \; (x,y,\lambda,N) \; \in \; \mbox{\rm
Capt}_{(\ref{eq:welldifcinclauxdis})}({\cal K},{\cal C}) \right\}
\end{displaymath}
where the subscript (\ref{eq:welldifcinclauxdis}) in ${\rm
Capt}_{(\ref{eq:welldifcinclauxdis})}({\cal K},{\cal C})$ refers
 to the discrete evolutionary system (\ref{eq:welldifcinclauxdis}). \color{Black}

The capture basin algorithm defines recursively a sequence of subsets
${\cal C}_{n}$ starting at ${\cal C}_{0}$ by

\begin{displaymath}
{\cal C} _{n+1} \; := \; {\cal K} \cap \left({\cal C}_{n} \cup \Psi  ^{-1}
({\cal C}_{n} ) \right)
\end{displaymath}
which converges to the capture basin  $\displaystyle{{\rm
Capt}_{(\ref{eq:welldifcinclauxdis})}({\cal K},{\cal C})}$.
 $\;\; \blacksquare$
\normalsize \vspace{ 5 mm}

 One can prove that whenever the discrete map $x  \leadsto \Phi
 _{h}(x):=x+hF_{h}(\lambda;x)$ is a time discretization of the
 differential inclusion $x'(t) \in F(\lambda;x(t))$, the graph of the
 discrete well converges to the graph of the well in the
Kuratowski-Painlev\'e sense (see \cite[Saint-Pierre]{ssp92vk,PSPAbcSHy}
and see \cite[Cardaliaguet, Quincampoix \& Saint-Pierre]{cqsp95dg} among
other references).

\color{ProcessBlue}  \section{Complete wells} \color{Black}

The concept of well we proposed in Definition~\ref{def:well} is not
restrictive enough   to match its physical counterpart \color{Black} in
the sense that it does not require all evolutions starting from a point of
a well ${\bf P}_{V}(\lambda;T,y)$ to remain below the rim of the well
before time $T$ while one of them at least reaches its rim at time $T$.

\vspace{ 5 mm} \fbox{\begin{minipage}{14.0cm} \begin{Minipage}
\label{mpdef:completewell}
\begin{center} \color{Red} {\large \bf Complete wells of a function under a
differential inclusion} \end{center}  \end{Minipage}
\footnotesize \color{PineGreen}
\begin{Definition}\label{def:completewell} Consider an  extended
function $ V:X \mapsto \mathbb{R}\cup \{+\infty\}$ and a differential
inclusion
\begin{displaymath} \forall \; t \geq 0, \;  \; x' (t) \; \in  \;
        F(\lambda;x(t))
 \end{displaymath}
 The \index{complete well of a function} \textsf{complete well}
${\bf W}_{V}(\lambda; T,y) \subset {\bf P}(V;y)$ of duration $T$
and depth $y $ of the function $V$ is defined by the set of initial
states $x \in {\bf S}(V;y)$ such that
\begin{enumerate}
\item  {\sf all} solutions $x_{\lambda}(\cdot) \in {\cal
S}_{\lambda}(x)$ satisfy

\begin{displaymath}
 \forall \; t \in [0,T], \; \; V (x_{\lambda}(t)) \; \leq \; y
\end{displaymath}
 \item  {\sf at least one}  solution $x_{\lambda}(\cdot) \in {\cal
S}_{\lambda}(x)$ satisfies

\begin{displaymath}
 V (x_{\lambda}(T)) \; = \; y\\
\end{displaymath}
\end{enumerate}
  \end{Definition}
\color{Black} \end{minipage}} \normalsize \vspace{5 mm}

The complete wells can be characterized in terms  of absorption and
capture basins (See for instance \cite[Aubin]{avt,a00capb,a00hjbi}) and
\cite[Aubin, Bayen, Bonneuil \& Saint-Pierre]{ab2sp00}).

\vspace{ 5 mm}\fbox{\begin{minipage}{14.0cm}
\begin{Minipage} \label{}
\begin{center}\color{Red} {\large\bf Viability characterization of
complete wells}\color{Black}\end{center}  \end{Minipage} \footnotesize
\begin{Proposition}
Consider an extended function $ V:X \mapsto \mathbb{R}\cup
\{+\infty\}$ and a differential inclusion
\begin{displaymath} \forall \; t \geq 0, \;  \; x' (t) \; \in  \;
        F(\lambda;x(t))
 \end{displaymath}
We associate with it the auxiliary system of differential inclusions
(\ref{eq:welldifcinclaux}). The constrained set ${\cal K}$ and the targets
${\cal C}$ are ${\cal D}$ are defined by

\begin{displaymath}
{\cal K} \; := \; {\cal E}p(V) \times \mathbb{R}_{+}\times \mathbb{R}_{+}
\; \; and \; \; {\cal C} \; := \; \mbox{\rm Graph}(V) \times
\mathbb{R}_{+} \times \{0\}
\end{displaymath}

and \begin{displaymath}  {\cal D} \; := \; {\cal E}p(V) \times
\mathbb{R}_{+} \times \{0\}
\end{displaymath}
Then

\begin{displaymath}
{\bf W}_{V}(\lambda; T,y) \; = \;  \left\{ x \in X \; \mbox{ such that} \;
(x,y,\lambda,T) \; \in \; \mbox{\rm
Capt}_{(\ref{eq:welldifcinclaux})}({\cal K},{\cal C}) \cap \mbox{\rm
Abs}_{(\ref{eq:welldifcinclaux})}({\cal K},{\cal D}) \right\}
\end{displaymath}
\color{Black}

\end{Proposition} \color{Black}\end{minipage}}
\normalsize \vspace{ 5 mm}

\footnotesize {\bf Proof} --- \hspace{ 2 mm} Indeed, to say that
$(x,y,\lambda,T) \; \in \; \mbox{\rm
Capt}_{(\ref{eq:welldifcinclaux})}({\cal K},{\cal C}) \cap \mbox{\rm
Abs}_{(\ref{eq:welldifcinclaux})}({\cal K},{\cal D})$ amounts to saying
that
\begin{enumerate}   \item
$(x,y,\lambda,T) \; \in \; \mbox{\rm
Capt}_{(\ref{eq:welldifcinclaux})}({\cal K},{\cal C})$, and thus, as we
have seen, that  $x \in {\bf P}_{V}(\lambda; T,y)$.
 \item
$(x,y,\lambda,T) \; \in \; \mbox{\rm
Abs}_{(\ref{eq:welldifcinclaux})}({\cal K},{\cal D})$ means that for all
evolutions $x_{\lambda}(\cdot) \in {\cal S}_{\lambda}(x)$, there exists  a
time $t^{\star} \geq 0$ such that the associated auxiliary evolutions $$t
\to (x(t), y(t); \lambda(t); \tau(t)) \; = \; (x(t), y,\lambda,
T-t)$$ starting from $(x,y,\lambda,T)$ at $t=0$ reaches the target ${\cal
D}$ at time  $t^{\star}$ while staying meanwhile  in ${\cal K}$~:
\begin{displaymath} \left\{ \begin{array}{ll}
(i) & (x(t^{\star}),y, \lambda,T-t^{\star}) \; \in \; {\cal D}\\
(ii) & \forall \; t \in [0,t^{\star}], \; \; (x(t),\lambda,y,T-t) \; \in
\; {\cal K}\\
\end{array} \right. \end{displaymath}
The first condition is equivalent to both equation $t^{\star}= T$ and
inequality $V(x(T)) \leq y$. The second equation means that for every $t
\in [0,T]$,  $V(x(t)) \leq y$.
 \end{enumerate}
 These are the two properties stating that
$x$ belongs to the well ${\bf W}_{V}(\lambda; T,y) $.
 $\;\; \blacksquare$
\normalsize \vspace{ 5 mm}

\section{Conclusions}

Our objective in this investigation was to build
 exploration dynamics of
a landscape $V$ associating with a bound $\lambda$ on the
velocities of the exploration mechanism,  a duration $T$
and  an altitude $y$:

\begin{enumerate}

\item the set ${\bf P}_{V}(\lambda;T,y)$ of initial
 states $x$ below  altitude $y$ from
which starts at least one evolution climbing the landscape in order to
  reach the altitude $y$ at exactly prescribed  time $T$; Altitude $y$
 might be either a reference level, thus providing access
to the depth of the well, or chosen among  the values of
local maxima of the landscape function, thus providing access to the height
of the barriers separating the well from the other ones;

\item  an underlying dynamical system governing the evolutions
climbing the wells up to their rims the velocities of the exploration
mechanism are consistently chosen among
\begin{displaymath}
x'(t) \; \in \;  -\frac{\partial {\bf P}_{V}(\lambda;T-t,y)}{\partial t}
\end{displaymath}
 Hence, the exploration mechanism is no
longer an external stochastic  modification of the gradient equation, but an
intrinsic  set-valued
method involving the time derivative of the well.

    \end{enumerate}

This dynamic description of landscape topography has then been
reformulated in the framework of viability theory, which  provides a
constructive algorithm to characterize quantitatively the landscape, built
as a intrinsic exploration mechanism of energy landscapes; this mechanism could
be either a perturbed gradient method or an universal mechanism
independent of the energy function. The more refined notion  of complete
well, introduced is Section~4, allows to bridge still more our
mathematical definitions and exploration with the current
landscape paradigm. As discussed in introduction (\S~1.2), our results can
then be exploited for taxonomic purposes, to investigate phase
transitions, to quantify the landscape hierarchical structure.
 It also proposes an alternative to standard
stochastic gradient methods,
 namely differential inclusions,
in modeling dynamics associated with an experimentally determined landscape.

\color{Black}

\begin{acknowledgments}
Jean-Pierre Aubin acknowledges the financial
support provided through the European Community's Human Potential
Programme under contract HPRNCT-2002-00281 (Evolution Equations for
Deterministic and Stochastic Systems).
\end{acknowledgments}

\newpage
\appendix

\color{ProcessBlue}  \section{Elements of Viability Theory} \color{Black}

Let $X$ be a finite-dimensional vector space.
A {\it set-valued map}  $F:X  \leadsto X$
associates to any $x\in X$ a subset $F(x)\subset X$.
The set-valued map $F$ generates
 the evolutionary system ${\cal
S}_F:X \leadsto {\cal C}(0,\infty;X)$ associating with any initial state $x_0
\in X$ the set ${\cal S}_F(x_0)\subset{\cal C}(0,\infty;X) $
 of solutions to {\it differential inclusion}
$x'(t)\in F(x(t))$ starting at $x_0$.
We denote by
\begin{displaymath}
 \mbox{\rm Graph}(F) := \{ (x,y) \in X \times Y \; | \; y \in F
(x)\} \subset X\times Y
\end{displaymath}
the {\it graph} of a set-valued map  $F:X \leadsto Y$ and $\mbox{\rm
Dom}(F) := \{x \in X | F (x) \ne \emptyset \}$ its {\it domain}.

We shall say that  a subset {\it $K \subset X$ is locally viable under
$F$\/}  (or under ${\cal S}_F$)
if from every $x \in K$ starts
 {\it at least one}
 solution $x ( \cdot )$ to the
differential inclusion $x' \in F (x)$ {\it viable in $K$} on the nonempty
interval $[0,T_{x}[$ in the sense
\begin{displaymath}
  \forall \;  t \in [0,T_{x}[, \;  \; x (t) \; \in K \;
\end{displaymath}
and that $K$ is {\it viable} if we can take $T_{x}=+ \infty $
for any $x\in K$.
Most of the results of viability theory are true whenever
we assume that the dynamics is Marchaud:

\vspace{ 5 mm} \fbox{\begin{minipage}{14.0cm} \begin{Minipage} \label{}
\begin{center} \color{Red} {\large \bf Marchaud maps} \end{center}  \end{Minipage}
\footnotesize \color{PineGreen} \begin{Definition} We shall say that
the set-valued map
 $F: X\leadsto Y $  is a {\sf Marchaud map}
\index{Marchaud map} if
        \begin{displaymath} \left\{ \begin{array}{ll}
        i) & \mbox{the graph of $F$  is closed in $X\times Y$}\\
         ii) & \mbox{the values $F (x)$ of $F$ are convex subsets of $Y$ }\\
        iii) & \mbox{the growth of F is linear:} \:\exists \; c>0
\; | \\
        &  \forall  x \in X, \; \|F (x)\|:=\sup_{v \in F (x)} \|v\|  \leq   c (
\|x\|+1)
        \end{array} \right. \end{displaymath}
        We shall say that $F$ is {\sf $ \lambda $-Lipschitz} if
(set-valued extension of the standard Lipschitz property)
        \begin{displaymath}
         \forall \; x,x' \in X, \;  \; F (x) \; \subset  \; F (x')+
\lambda \|x-x'\|B
        \end{displaymath}
        where $B$ is the unit ball in $Y$.\end{Definition}
\color{Black} \end{minipage}} \normalsize \vspace{5 mm}

        We shall also need some other prerequisites from
Viability Theory:

\vspace{ 5 mm} \fbox{\begin{minipage}{14.0cm} \begin{Minipage}
\label{mpdef:capturebasin}
\begin{center} \color{Red} {\large \bf Capture and absorption basins} \end{center}  \end{Minipage}
\footnotesize \color{PineGreen} \begin{Definition}
\label{def:capturebasin}
        Let  $C \subset K \subset  X$ be two subsets, $C$ being
regarded as a target, $K$ as a constrained set.  The subset $ \mbox{\rm
Capt}_F(K,C)$ of initial states $x_{0} \in K$ such that $C$ is reached in
finite time,
 without leaving $K$, \color{PineGreen}
  by {\sf at least one} solution $x(\cdot
) \in  {\cal S}_F (x_{0})$  starting at $x_{0}$
is called the {\sf
viable-capture basin of $C$ in $K$\/} \index{viable-capture basin}
\color{Black}
(the solution might eventually leaves $K$, but only after
 having reached $C$).
\color{PineGreen}
The subset $ \mbox{\rm Abs}_F(K,C)$ of initial states $x_{0} \in K$ such
that \textbf{all  evolutions} $x(\cdot ) \in  {\cal S}_F (x_{0})$ starting
at $x_{0}$ are  viable in $K$ until they reach $C$ in finite time  is
called the {\sf absorption basin of $K$ with target $C$\/}.
\index{absorption basin of $K$ with  target $C$}
        \end{Definition} \color{Black}
\end{minipage}} \normalsize \vspace{5 mm}

 Obviously $\mbox{\rm Abs}_F(K,C)\subset \mbox{\rm
Capt}_F(K,C)$.
We recall the following result \cite[Aubin~\&~Catt\'e]{ac01vka}:

\vspace{ 5 mm} \fbox{\begin{minipage}{14.0cm} \begin{Minipage}
\label{mpdetlargcaptbasprp0}
\begin{center} \color{Red} {\large \bf Bilateral fixed point property} \end{center}  \end{Minipage}
\footnotesize \color{Orange}\begin{Theorem} \label{detlargcaptbasprp0}

The
viable \color{Orange}-capture basin $ \mbox{\rm Capt}_{}(K,C)$
of a target $C$
viability being with respect to the constrained
set  $K$)
\color{Orange} is  the {\sf unique} subset $D $ satisfying $C \subset  D \subset K$ and

\begin{displaymath}
 D \; = \; \mbox{\rm Capt}_{F}(K,D) \; = \; \mbox{\rm
Capt}_{F}(D,C)
\end{displaymath}
and the absorption basin of $K$ with target $C$ is  the {\sf unique}
subset $A $ satisfying $C \subset  A \subset K$ and
\begin{displaymath}
 A \; = \; \mbox{\rm Abs}_{F}(K,D) \; = \; \mbox{\rm
Abs}_{F}(D,C)
\end{displaymath}
 \end{Theorem} \color{Black}
\end{minipage}}\normalsize \vspace{ 5 mm}

\vspace{ 5 mm} \fbox{\begin{minipage}{14.0cm} \begin{Minipage} \label{}
\begin{center} \color{Red} {\large \bf Backward invariance} \end{center}  \end{Minipage}
\footnotesize \color{PineGreen} \begin{Definition}

The subset $K$ is {\sf
locally backward invariant under $F$\/} if for every $t_{0} \in ]0, +
\infty [,  \;  x \in  K$, for {\sf  all}
 solutions \index{backward viable set
under a set-valued map}\index{invariant set under a set-valued map} $x(
\cdot )$ to the differential inclusion $x' \in  F  (x)$  arriving at $x$
at time $t_{0}$, there exists a time $s \in [0,t_{0}[$
(depending on the solution)  such that $x ( \cdot )$
is viable in $K$ on the interval $[s,t_{0}]$. The subset $K$
is  {\sf backward invariant} under $F$
if we can take $s=0$ for all solutions.
\end{Definition}
\color{Black} \end{minipage}} \normalsize \vspace{5 mm}

It is straightforward to check that backward evolutions
$\theta\to z(\theta)=x(t_0-\theta)$ are solutions
 of the differential inclusion $z'(\theta)\in -F(z(\theta))$
with initial condition $z(0)=x_0$ if $x(t_0)=x_0$;
we call them {\it backward solutions} (starting from
$x_0$ at time $\theta=0$).
It is to note that (local) backward invariance of $K$
is stronger than (local) viability of $K$ under this backward
evolution, since {\it all} solutions starting in
a backward invariant subset $K$ remain
in $K$ for a finite time (depending on each considered solution
in case of the local version of the property),
whereas (local) viability of $K$ only requires that for each point
$x\in K$, at least {\it one} solution is (locally) viable in $K$.

We also introduce a weaker notion:
 A subset $D\subset K$ is {\it locally backward invariant
relatively to} $K$ if all backward solutions starting from $D$
and  viable in
$K$ (i.e. remaining in $K$ for a finite time)
are actually viable in $D$ (i.e. remain in $D$ for a finite time).

\vspace{ 5 mm} \fbox{\begin{minipage}{14.0cm} \begin{Minipage} \label{}
\begin{center} \color{Red} {\large \bf Repellers} \end{center}  \end{Minipage}
\footnotesize \color{PineGreen} \begin{Definition}
 A subset $R\subset X$ is a {\sf repeller} under $ F$
\index{repeller} if all solutions starting from $R$ leave $R$
 in finite
time.
\end{Definition}
\color{Black} \end{minipage}} \normalsize \vspace{5 mm}

 Hence, $R$ is not viable, but this does not exclude
local viability. It is moreover obvious that any subset of a repeller is itself
a repeller.

We can derive the following characterization of capture basin (see
\cite[Aubin]{a00capb}):

\vspace{ 5 mm} \fbox{\begin{minipage}{14.0cm} \begin{Minipage} \label{}
\begin{center} \color{Red} {\large \bf
Characterization of capture basins} \end{center}  \end{Minipage}
\footnotesize \color{Orange}\begin{Theorem} \label{captvkthm002} Let us
assume that  $ F$ is Marchaud and that the subsets $C \subset K$ and $K$
are closed. If $K \backslash C$ is a repeller (this is
for instance  the case when $K$
itself is a repeller), then the viable-capture basin $ \mbox{\rm
Capt}_F(K,C)$ of the target $C$ under $F$ is the {\sf unique} closed
subset $D$  satisfying $C \subset  D \subset K$ and \footnote{The subset $K
\backslash C$ denotes the intersection of $K$ and the complement of $C$,
i.e., is the set of elements of $K$ which do not belong to $C$.}
\begin{displaymath} \left\{ \begin{array}{ll}
i) & \mbox{\rm $D \backslash C$ is {\sf locally viable} under $F$} \\
ii) & \mbox{$D$ is locally backward invariant relatively to
$K$ }\\
\end{array} \right. \end{displaymath} \end{Theorem} \color{Black}
\end{minipage}}\normalsize \vspace{ 5 mm}

\vspace{ 5 mm} \fbox{\begin{minipage}{14.0cm} \begin{Minipage} \label{}
\begin{center} \color{Red} {\large \bf Contingent cones} \end{center}  \end{Minipage}
\footnotesize \color{PineGreen}
\begin{Definition}
The contingent cone $T_{L} (x)$ to $L \subset X$ at $x \in L$ is the set
(obviously a closed cone)
of directions $v \in X$ such that there exist sequences $h_{n}
>0$ converging to $0$ and $v_{n}$ converging to $v$
satisfying
$x+h_{n}v_{n} \in L$
\color{PineGreen}  for every $n$  (see for instance \cite[Aubin \&
Frankowska]{af90sva}).
\end{Definition} \color{Black}
\end{minipage}}\normalsize \vspace{ 5 mm}

For instance, if $L$ is a differentiable manifold in $X$,
$T_L(x)$ coincide with the tangent space to $L$ at point $x$.
If the interior of $L$ is non-empty, then $T_L(x)=X$ for
any $x\in{\rm Int}(L)$.

 We introduce the following Frankowska property that
we need for deriving the system of Hamilton-Jacobi-Bellman equations of
which the well is a solution:

\vspace{ 5 mm} \fbox{\begin{minipage}{14.0cm} \begin{Minipage} \label{}
\begin{center} \color{Red} {\large \bf The Frankowska property} \end{center}  \end{Minipage}
\footnotesize \color{PineGreen} \begin{Definition} Let us consider a
set-valued map $F:X \leadsto X$ and two subsets $C \subset K$ and $K$. We
shall say that a subset $D$ between $C$ and $K$
(i.e. $C\subset D\subset K$) satisfies the {\sf
Frankowska property} \index{Frankowska property} with respect to $F$ if

\begin{equation} \left\{ \begin{array}{ll} \label{franprtgeq}
i) & \forall \; x \in D \backslash C, \; F (x) \cap T_{D} (x)  \ne
\emptyset \\
ii) & \forall \; x \in D  \;\mbox{\rm such that \footnote{This is always
satisfied when $x \in \mbox{\rm Int}(K)$. }}\; \; -F (x)
\cap   T_{K} (x)  \ne  \emptyset\\
& \;\mbox{\rm then}\; -F (x)  \subset   T_{D} (x)
\end{array} \right. \end{equation}
When $K$ is assumed further to be locally backward
invariant
(then $ -F (x)\subset  T_{K} (x)$ for any $x\in K$)
\color{PineGreen} the above
conditions (\ref{franprtgeq})  boil down to
\begin{equation} \left\{ \begin{array}{ll}
\label{chacaptbasineq0346} i) & \forall \; x \in D \backslash C, \; \;F
(x) \cap T_{D} (x)
 \ne
\emptyset \\
ii) &  \forall \; x \in D, \;  \; -F (x) \subset  T_{D} (x)\\
\end{array} \right. \end{equation} \end{Definition}
\color{Black} \end{minipage}} \normalsize \vspace{5 mm}

(The minus sign in front of $F$ arises when considering
backward evolution, governed by the differential inclusion
$z'(\theta)\in -F(z(\theta))$.)

Theorem~\ref{captvkthm002} and the Viability \footnote{See  for instance
Theorems~3.2.4,  3.3.2  and 3.5.2  of   \cite[Aubin]{avt}.} and Invariance
Theorems imply

\vspace{ 5 mm} \fbox{\begin{minipage}{14.0cm} \begin{Minipage} \label{}
\begin{center} \color{Red} {\large \bf Viability characterization
of capture basins} \end{center}  \end{Minipage}
\footnotesize \color{Orange}\begin{Theorem} \label{viablcaptbascharthm1}
Let us assume that $F$ is Marchaud, that $K$  and $C \subset K$
are closed subsets and that $K \backslash C$ is a repeller. Then  the
capture basin $ \mbox{\rm Capt}_{F}(K, C)$ is
\begin{enumerate}
\item the {\sf largest}  closed subset $D$ satisfying  $C \subset
D \subset K$ and
\begin{equation} \label{wonderfulthmeq27}
  \forall \; x\in D \backslash C, \; \; F (x) \cap T_{D} (x) \;
\ne \; \emptyset
 \end{equation}
Furthermore, the evolutions $x(\cdot) \in  {\cal S}_F(x)$ viable in $K$
until they reach $C$ are governed by the differential inclusion

\begin{displaymath}
x'(t) \; \in \;  F (x(t)) \cap T_{D} (x(t))
\end{displaymath}
(It roughly means that these trajectories point into
$D$ at any point where they reach the boundary of $D$, thus ensuring
their viability until they reach $C$.)

\color{Orange}
\item  if $F$ is Lipschitz, the {\sf unique}  closed subset $D$
satisfying the  Frankowska property (\ref{franprtgeq}).

\noindent
The absorption basin $ \mbox{\rm Abs}_F (K,C)$ is
 the {\sf largest}  closed subset $D$ satisfying  $C \subset
D \subset K$ and
\begin{equation} \label{wonderfulthmeq27}
  \forall \; x\in D \backslash C, \; \; F (x) \subset  T_{D} (x)
 \end{equation}
\end{enumerate}\end{Theorem} \color{Black}
\end{minipage}}\normalsize \vspace{ 5 mm}

We shall apply Theorem~\ref{viablcaptbascharthm1} to the case when subsets
$K := \mbox{\rm Graph}(F)$, $C:= \mbox{\rm Graph}(H)$ are graph of
set-valued maps from $X$ to $X$ and when we decide to regard $D$ as the
graph of a set-valued map $G: \mathbb{R} \times X  \leadsto Y$. We
then interpret the contingent cone to the graph as the graph of the
contingent derivative.  We  obtain set-valued solutions to systems of
Hamilton-Jacobi inclusions
that this unknown function $G$ should satisfy in order that its graph
yields the desired capture  basin.
We refer to
\cite[Aubin]{a00hjbi,ab2sp00}, \cite[Aubin \& Frankowska]{af91hyp} and
their references for more details on this topic. Here, we recall the
definition of contingent derivative of a set-valued map and translate
Theorem~\ref{viablcaptbascharthm1} in the framework of wells.

\vspace{ 5 mm} \fbox{\begin{minipage}{14.0cm} \begin{Minipage} \label{}
\begin{center} \color{Red} {\large \bf Contingent derivative
 of a set-valued map} \end{center}  \end{Minipage}
\footnotesize \color{PineGreen} \begin{Definition}
\label{def:contingentderivative} Let us consider a set-valued map $G:
\mathbb{R}^{} \times X  \leadsto Y$. The graph  of the contingent
derivative $DG(t,x,y)$ (a set-valued map defined
from $ {\bf R} \times X$ to $Y$) at a point $(t,x,y)
\in \mbox{\rm Graph}(G)$ is equal to the contingent cone to the graph of
$G$ at $ (t,x,y)$:
 \begin{displaymath}
         T_{ \mbox{\rm Graph}(G)} (t,x,y) \; = \; \mbox{\rm
Graph}(DG (t,x,y))
        \end{displaymath}
\end{Definition}
\color{Black} \end{minipage}} \normalsize \vspace{5 mm}

Consequently, to say that $w\in Y$ belongs to the contingent derivative
$DG(t,x,y) ( \pm 1,v)$ of $G$ at $ (t,x,y)$ in the direction $( \pm 1,v)
\in {\bf R} \times X$ means that
\begin{displaymath}
 \liminf_{h \rightarrow 0+, \; v' \rightarrow v}d \left( w, \frac{G
(t \pm h,x+hv')-y}{h}  \right) \; = \; 0
\end{displaymath}
where $d$ is any distance in $Y$.
Since the contingent cone is a closed subset, the graph of a contingent
derivative is always closed and positively homogeneous (this is what
remains of the required linearity of the derivative in classical analysis,
but, fortunately, we can survive pretty well without linearity).

When $g: {\bf R} \times X \mapsto Y$ is single-valued, we set $Dg
(t,x):=Dg (t,x,g (t,x))$. We see at once that $Dg (t,x) ( \pm 1, v)= \pm
\frac{ \partial g (t,x)}{ \partial t}+ \frac{ \partial g (t,x)}{ \partial
x} \cdot v$ whenever $g$ is differentiable at $ (t,x)$. The above definition (\ref{def:contingentderivative}) generalizes
to set-valued maps a property obviously valid for differentiable maps,
hence provides a consistent extension of the differentiation  to
 set-valued maps,  coinciding with the plain notion
 for smooth single-valued maps.
Moreover, it is to  note that
when $g$ is
Lipschitz on a neighborhood of $ (t,x)$ and when the dimension of $X$ is
finite, the domain of $Dg (t,x)$ is not empty. Furthermore, the Rademacher
Theorem stating that a locally Lipschitz single-valued map is almost
everywhere differentiable implies that $x \leadsto Dg (t,x)$ is almost
everywhere single-valued.
However, in this case, equality $Dg (t,x) (-1,-v)=-Dg (t,x) (1,v)$ is not
true in general. We refer to \cite[Aubin \& Frankowska]{af90sva} for more
details.

\footnotesize {\bf Remark:} --- \hspace{ 2 mm}   This is how Fermat
defined in 1637 the derivative of a function as the slope of the tangent
to its graph. Leibniz and Newton provided the characterization in terms of
limits of difference quotients. Here, too,
          {\em the graph of the contingent derivative $DG (t,x,y)$ is the
upper Painlev\'e-Kuratowski  limit of the graphs of difference
quotients\/}
$\nabla_{h}G(t,x,y)$  of $G$ at
$(t,x,y) \in \mbox{\rm Graph}(G)$ defined by
        $$ ( \lambda ,v) \mapsto  \nabla_{h}G (t,x,y) ( \lambda
,v) \; := \;  \frac{G ( t+ \lambda  h, x+hv)-y}{h}$$
        Indeed, we observe that
        \begin{displaymath}
         \mbox{\rm Graph}(\nabla_{h}G (t,x,y)) \; = \; \frac{
\mbox{\rm Graph}(G)- (t,x,y)}{h} \;\;(\subset {\bf R}\times X\times Y)
        \end{displaymath}
        so that the contingent cone to the graph of $G$, being the
upper limit of the graphs of the difference quotients, is equal by
definition to  the graph of the upper graphical limit of the difference
quotients.

The strong requirement of pointwise convergence of differential quotients
involved in the usual derivatives
can be  weakened  in  (at least)  two ways,  each way sacrificing
different groups of properties of these usual derivatives:

\begin{itemize}
\item {\bf Distributional Derivatives}:  Fix the direction $v$ and take the
limit of the function $x \mapsto \nabla _{h}g (x)  (v)$ in the weaker
sense of  distributions.  The limit $D_{v}g$  may then be a distribution,
and no longer a single-value map.  However, it coincides with the usual
limit
($D_vg(x)=Dg(x).v$)
 when $g$  is  G\^ateaux  differentiable.  Moreover,  one  can define
difference quotients of  distributions,  take  their  limit,  and thus,
differentiate distributions.

Distributions are no longer functions or maps defined on $  {\bf  R}^{n}
$, so  these distributional derivatives
 loose  the pointwise character of  functions and maps;
 on the other hand, this generalization
retains the linearity of the operator $g \mapsto D_{v}g$,  mandatory for
using the theory of linear operator for solving partial differential
equations.

\item {\bf Graphical Derivatives}: Fix the point $x$ and take the limit
of the  function $v \mapsto \nabla  _{h}g (x)  (v)$  in the weaker sense
of graphical  convergence  (the  graph  of  the  graphical  limit being by definition
the Painlev\'e-Kuratowski upper limit of  the graphs).  The limit $Dg (x)$
may then  be a set-valued map,  and no  longer a single-value map.
However, it coincides with  the  usual  limit  when  $g$  is  G\^ateaux
differentiable. Moreover,  one  can define  difference  quotients of
set-valued maps, take their
graphical  limit, and thus, differentiate set-valued
maps. These graphical derivatives keep  the pointwise character of
functions and maps,  mandatory for implementing the Fermat Rule, proving
inverse function theorems under constraints or  using Lyapunov functions,
for instance, but loose the linearity of the map $g \mapsto Dg (x)$.
\end{itemize}

   In  both  cases,   the  approaches  are  similar:  They  use (different)
convergences {\em weaker than the pointwise convergence}  for increasing
the possibility for the  difference-quotients  to  converge. But the price
to pay is the loss of some properties by passing to these weaker limits
(the pointwise character  for distributional derivatives,  the linearity
of  the  differential  operator  for graphical derivatives). $\;\;
\blacksquare$ \normalsize \vspace{ 5 mm}

Proposition~\ref{prp:viabcahrwell} related the graph of the well to the
capture basin

\begin{displaymath}
{\bf P}_{V}(\lambda; T,y) \; = \;  \left\{ x \in X \; \mbox{ such that} \;
(x,y,\lambda,T) \; \in \; \mbox{\rm
Capt}_{(\ref{eq:welldifcinclaux})}({\cal K},{\cal C}) \right\}
\end{displaymath}
under system (\ref{eq:welldifcinclaux}):

\begin{displaymath} \begin{array}{lcr} \left\{ \begin{array}{ll}
 (i) & x'(t) \; \in \; F(\lambda(t);x(t))\\
(ii) & y'(t) \; = \; 0\\
 (iii) & \lambda'(t) \; = \; 0\\
 (iv) & \tau'(t) \; = \; -1\\
 \end{array} \right.&&\hskip 15mm (1)\end{array} \end{displaymath}

\noindent
 At this point, we need to introduce the concepts of epigraph and
 epiderivative of extended numerical functions:

\vspace{ 5 mm} \fbox{\begin{minipage}{14.0cm}
        \begin{Minipage} \label{mpEpigraph} \begin{center}
       \color{Red} {\large \bf Epigraph and epiderivative of a function}
        \end{center}  \end{Minipage} \footnotesize
     \color{PineGreen}
       \begin{Definition} \label{defEpigraph}
Let  $V:X \mapsto \mathbb{R}\cup \{+\infty \}$ be an extended function.
Its \hypertarget{epigraph}{\textsf{epigraph}} \index{epigraph}
     $\mathcal{E}p(V)$  is the set of pairs
     $(x,y)\in  X \times \mathbb{R}$ satisfying
     $V(x) \leq y$
(thus $\mathcal{E}p(V)\subset X\times {\bf R}$).
\color{PineGreen}
The \textsf{contingent epiderivative} $D_{\uparrow}V(x): X \mapsto
\overline{\mathbb{R}}$ is defined through the relation
\begin{displaymath}
\mathcal{E}p(D_{\uparrow}V(x)) \; := \; T_{\mathcal{E}p(V)}(x,V(x))
\end{displaymath}
         \end{Definition} \color{Black}
        \end{minipage}}\vspace{ 5 mm}

        We can check that
 $D_{\uparrow}V(x)$ consistently coincide with
the usual derivative $DV(x)$ when $V$ is differentiable in $x$ and that
 for any $v \in X$,

\begin{displaymath}
D_{\uparrow}V(x)(v) \; = \; \liminf_{h  \rightarrow 0+,\; v' \rightarrow
v} \frac{V(x+hv')-V(x)}{h}
\end{displaymath}
is a generalized limit of differential quotients.

 We deduce from
Proposition~\ref{prp:viabcahrwell} and Theorem~\ref{viablcaptbascharthm1}
the following characterization of the well as the unique solution to an
initial-value problem of a partial differential inclusion satisfying
viability constraints:

\vspace{ 5 mm} \fbox{\begin{minipage}{14.0cm} \begin{Minipage}
\label{mpthm:pdiwell}
\begin{center} \color{Red} {\large \bf
Intrinsic exploration mechanism and the well partial differential inclusion}
\end{center}  \end{Minipage} \footnotesize \color{Orange}\begin{Theorem}
\label{thm:pdiwell} Assume that the set-valued map $F$ is Marchaud and
that the function $V$ is continuous. Then the well ${\bf P}_{V}:
\mathbb{R}_{+} \times \mathbb{R}_{+} \times \mathbb{R}  \leadsto  X$
is the largest set-valued map ${\bf P}: \mathbb{R}_{+} \times
\mathbb{R}_{+} \times \mathbb{R}  \leadsto X$ solution to the partial
differential inclusion
\begin{displaymath}
\forall \; x \in {\bf P}(\lambda; T,y), \; \; F(\lambda;x)
 \cap D {\bf P}(\lambda; T,y,x)(0,-1,0) \ne \emptyset
\end{displaymath}
the initial condition $${\bf P}(\lambda; 0,y)={\bf S}_{0} (V, y)$$ and
the viability constraint
$${\bf P}(\lambda; T,y) \subset {\bf S} (V, y)$$

Furthermore, if $F$ is Lipschitz, this solution is the unique solution
satisfying
\begin{displaymath} \left\{ \begin{array}{ll}
(i) & \forall \; x \in {\bf P}(\lambda; T,y), \; \; F(\lambda;x)
 \cap D {\bf P}(\lambda; T,y,x)(0,-1,0) \ne \emptyset\\
(ii) &  \forall \; x \in {\bf P}(\lambda; T,y) \;\mbox{\rm such
that}\; \inf _{v\in F(\lambda;x)} D_{\uparrow}V(x)(-v) \leq 0. \\
 &\;\mbox{\rm then}\; -F (\lambda;x) \subset D {\bf P}(\lambda; T,y,x)(0,+1,0)
\end{array} \right.
\end{displaymath}
\end{Theorem} \color{Black}
\end{minipage}}\normalsize \vspace{ 5 mm}

\footnotesize {\bf Proof} --- \hspace{ 2 mm}
 Theorem~\ref{viablcaptbascharthm1} implies that
the graph of  the well ${\bf P}_{V}: \mathbb{R}_{+} \times \mathbb{R}_{+}
\times \mathbb{R}  \leadsto  X$,
once transformed by the permutation
$(\lambda, \tau, y,x)\to (x,y, \lambda, \tau)$
of the coordinates
 is the largest subset ${\cal D}$
between ${\cal C}$ and ${\cal K}$
(i.e. $ {\cal C} \subset {\cal D}
\subset {\cal K} \subset X\times \mathbb{R}
\times \mathbb{R}_{+} \times \mathbb{R}_{+}$)
 such that

\begin{displaymath}
\forall \; (x,y,\lambda,\tau), \; \; (F(\lambda;x) \times \{0\} \times
\{0\}\times \{-1\}) \; \cap \; T _{{\cal D}} (x,y,\lambda,\tau) \ne
\emptyset
\end{displaymath}
This amounts to saying that the well ${\bf P}_{V}$ is the largest
set-valued map ${\bf P}$ satisfying the initial condition ${\bf
P}(\lambda; 0,y)={\bf S}_{0} (V, y)$, the constraint ${\bf
P}(\lambda; T,y) \subset {\bf S} (V, y)$ and the contingent solution
to the partial differential inclusion
\begin{displaymath}
\forall \; x \in {\bf P}(\lambda; T,y), \; \; F(\lambda;x)
 \cap D {\bf P}(\lambda; T,y,x)(0,-1,0) \ne \emptyset
\end{displaymath}
and that the evolutions $(t \mapsto (\lambda, T-t,y, x(t)))$ viable in the
well until they reach its rim are governed by the differential inclusion

\begin{displaymath}
(0,-1,0,x'(t)) \; \in \; (\{0\} \times \{-1\} \times \{0\} \times
F(\lambda;x(t))) \cap \mbox{\rm Graph}(D{\bf P}_{V}(\lambda;
T-t,y,x(t))(0,-1,0))
\end{displaymath}
This can be written

\begin{displaymath}
x'(t) \; \in \; F(\lambda;x(t)) \cap D{\bf P}_{V}(\lambda;
T-t,y,x(t))(0,-1,0)
\end{displaymath}
This is what we meant symbolically above as
\begin{displaymath}
x'(t) \; \in \; F(\lambda;x(t)) \cap -\frac{\partial {\bf
P}_{V}(\lambda;T-t,y)}{\partial t}
\end{displaymath}

When $F$ is Lipschitz (this is the case when $F(\lambda;x):= \lambda B$
 (where $B$ is the unit ball in $X$)
the graph of the well ${\bf P}_{V}$
(after permutation of the coordinates as above)
is the unique subset ${\cal D}$ satisfying
\begin{displaymath}
\forall \; (x,y,\lambda,\tau), \; \; (F(\lambda;x) \times \{0\} \times
\{0\}\times \{-1\}) \; \cap \; T _{{\cal D}} (x,y,\lambda,\tau) \ne
\emptyset
\end{displaymath}
and, whenever $(-F(\lambda;x) \times \{0\} \times \{0\}\times \{+1\}) \;
\cap \; T _{{\cal K}} (x,y,\lambda,\tau) \ne \emptyset$, then
\begin{displaymath}
(-F(\lambda;x) \times \{0\} \times \{0\}\times \{+1\}) \; \subset \; T
_{{\cal D}} (x,y,\lambda,\tau)
\end{displaymath}
Thanks to the definition of the contingent epiderivative and the fact that
${\cal K} \; := \; {\cal E}p(V) \times \mathbb{R}_{+}\times
\mathbb{R}_{+}$, we infer that

\begin{displaymath}
(-v,0,0, +1) \; \in \; T _{{\cal D}} (x,y,\lambda,\tau)
\end{displaymath}
if and only if  $D_{\uparrow}V(x)(-v) \leq 0$. This concludes the proof.
$\;\; \blacksquare$ \normalsize \vspace{ 5 mm}

\end{document}